\pdfoutput=1

\documentclass{aa}
\usepackage{graphicx}
\usepackage{txfonts}
\usepackage{natbib}
\usepackage{xspace}
\usepackage{hyperref}

\bibpunct{(}{)}{;}{a}{}{,} 

\begin{document}
%

\def\teff{T$_{e\! f\! f}$~}
\def\as{\arcsec\xspace}

\title{Exoplanet characterization with long slit spectroscopy} \subtitle{}

\author{Arthur Vigan \and Maud Langlois \and Claire Moutou \and Kjetil Dohlen}

\institute{Laboratoire d'Astrophysique de Marseille, Technop\^ole Marseille   \'Etoile, 38, rue Fr\'ed\'eric Joliot Curie
  13013 Marseille, France\\
  \email{arthur.vigan@oamp.fr} }

\date{Received 29 April 2008 / Accepted 22 July 2008}

 \abstract
{Extrasolar planets observation and characterization by high contrast imaging
  instruments is set to be a very important subject in observational astronomy. Dedicated instruments are being developed to achieve this goal with very high efficiency. In particular, full   spectroscopic characterization of low temperature planetary companions is an   extremely important milestone.}
{We present a new data analysis method for long slit spectroscopy (LSS) with   coronagraphy, which allows characterization of planetary companions of low   effective temperature. In a speckle-limited regime, this method allows an   accurate estimation and subtraction of the scattered starlight, to extract a clean spectrum of the planetary companion.  }
{We performed intensive LSS simulations with IDL/CAOS to obtain realistic   spectra of low ($R=35$) and medium ($R=400$) resolution in the J, H, and K   bands. The simulated spectra were used to test our method and estimate   its performance in terms of contrast reduction and extracted spectra   quality. Our simulations are based on a software package dedicated to the development of SPHERE, a second generation instrument for the ESO-VLT.}
{Our method allows a contrast reduction of 0.5 to 2.0 magnitudes compared to the   coronagraphic observations. For M0 and G0 stars located at 10~pc, we show that   it would lead to the characterization of companions with \teff of 600~K and   900~K respectively, at angular separations of 1.0\as. We also show that errors   in the wavelength calibration can produce significant errors in the   characterization, and must therefore be minimized as much as possible.}
{}


 \maketitle


\section{Introduction}
\label{section:introduction}

Detection of exoplanets through indirect methods, such as radial velocities or transits has become a subject of intensive research that has led to the discovery of more than 270 exoplanets. A future generation of instruments will be able to deliver very high contrast coronagraphic images of nearby stars to detect possible orbiting companions. These direct detections are particularly effective at detecting relatively hot companions at large separations from their host stars, making them the perfect complement to radial velocity and transit methods that are usually able to detect planets with short orbiting periods.

The European project SPHERE (Spectro-Polarimetic High contrast imager for Exoplanets REsearch) for the VLT second generation of instruments aims to detect exoplanets down to a contrast of $10^{-6}$ at angular separations as small as 0.1\as with dedicated extreme adaptive optics and coronagraphs \citep{dohlen2006}. One of the 3 science channels of SPHERE will be equipped with a near infrared camera, IRDIS (InfraRed Dual Imaging Spectrograph), that will offer several observing modes for exoplanet detection and characterization including long slit spectroscopy (LSS). The prime objective of IRDIS is exoplanet direct detection with simultaneous differential imaging (SDI), a technique that has been extensively described in the literature \citep{racine1999,marois2000} including possible improvements \citep{marois2006}. It was also used successfully to detect cool companions with the NACO instrument on the ESO-VLT \citep{lenzen2004}. This technique relies on the fact that the planetary spectrum has deep absorption features due to its atmosphere composition (e.g. CH$_{4}$ in the H band), whereas the starlight has a relatively smooth spectrum over the observation band. Recording two simultaneous images inside and outside of the planet absorption band, and subtracting them allows removal of most of the starlight, leaving mainly the planetary light in the subtracted image residuals.

Although very powerful at detection, this technique appears less suitable for the characterization of exoplanets. Current models predict that giant gas planet atmospheres most probably contain methane and water \citep{allard2001,allard2003,burrows1997,burrows2006}. The reliance on the related molecular features for detection is a primary step in characterization, since it identifies planetary objects with atmosphere deprived of these compounds. However, a full characterization is not possible with a flux difference of only 2 narrow and adjacent bands, which is the information provided by the SDI technique: some degeneracy in planet parameters is probable if only the methane band in H is observed. It can also be difficult to exclude signal contamination by a background object without more information. This explains why follow-up spectroscopic observations over a significantly wide wavelength range are required.

In IRDIS, this observation mode is achieved with long slit spectroscopy at small ($R=35$) and medium ($R=400$) resolutions and coronagraphy over J, H and K bands. In this context, we adapted and improved the data reduction technique first developed by \citet{sparks2002} and used by \citet{thatte2007} with AO-fed Integral Field Spectroscopy (IFS), to extract a full companion spectrum from simulated high-contrast images. In contrast to its use with an IFS, LSS provides spatial information in only one dimension (along the slit), making it necessary to develop a specific data analysis method.

Our method is very general and can be used on coronagraphic images as well as non-coronagraphic ones. Moreover, it is fairly simple to implement, and does not require a large number of operations. It is fully described here in the context of SPHERE, since it is intended to be used for LSS data reduction on this particular instrument. Section \ref{section:data_analysis} describes the concept and practical implementation details, Sect. \ref{section:lss_simulations} provides a global overview of the simulations performed to simulate realistic LSS data, and Sect. \ref{section:results} details the global performance of the method and the influence of a few key parameters.

\section{Data analysis of coronagraphic long slit spectroscopy}
\label{section:data_analysis}

\subsection{Speckles in long slit spectroscopy images}
\label{section:speckles_in_long_slit_sectroscopy_images}

\begin{figure}[t]
  \centering
  \includegraphics[width=8cm]{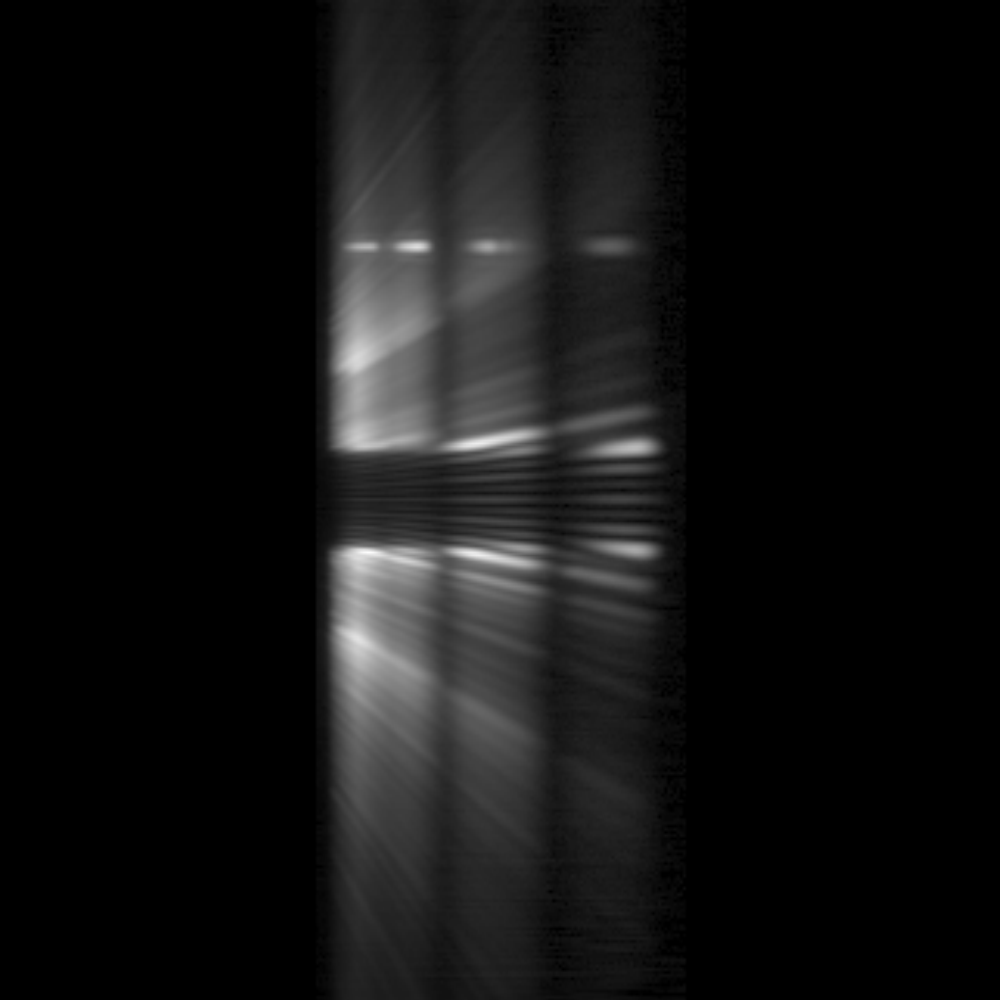}
  \caption{Low resolution ($R=35$) simulated spectrum with coronagraphy in the J, H and K bands of an M0 star with a relatively bright companion at an angular separation of 1.0\as. Wavelength is along the horizontal axis, and angular separation along the vertical axis, starting from the center. The obscured central part corresponds to the position of the 0.4\as coronagraph mask. The speckles form oblique lines in the spectrum because their angular separation scales linearly with wavelength, whereas the planet position is fixed. For clarity, the displaying scale is not linear.}
  \label{figure:typical_spectrum}
\end{figure}

The main limitation in high-contrast coronagraphic images originates in the speckles \citep{soummer2007} induced by atmospheric phase residuals and instrumental static and quasi-static aberrations not corrected by adaptive optics (AO). Most of the atmospheric residuals average out with time, producing a smooth halo on which the static and quasi-static speckles (sometimes called \emph{super-speckles} in the literature) are superimposed. The coherence time of the super-speckles extends from tens of seconds to several minutes \citep{macintosh2005,hinkley2007}, so it is safe to assume that they do not average during a typical integration. In the case of a single image at a given wavelength, the typical angular size of a speckle is $\theta_{speckle}=\lambda/D$, where $\lambda$ is the wavelength of observation and $D$ is the telescope diameter. The planet is assumed to be an unresolved point source and its image has a full width at half-maximum (FWHM) that is approximately equal to the diffraction limit. Differentiating a faint planet point-spread function (PSF) from a speckle is therefore impossible with a single image, and without using a temporal sequence or other spectral information.

The position and size of a speckle in the field of view is wavelength-dependent: as the wavelength increases from $\lambda_{1}$ to $\lambda_{2}$, the FWHM of a single speckle and its angular separation from the star both increase by a factor $\lambda_{2}/\lambda_{1}$. Following the wavelength axis at a fixed angular separation from the star, one would measure a strong modulation in the intensity as the Airy rings and speckles cross this position \citep[see][Fig.~24]{sparks2002}. A possible faint companion signal would remain undetected in the modulations due to the wavelength dependence of the speckle pattern. However, a fixed physical object (e.g. a planetary companion) will not change its position with wavelength: only its FWHM will be multiplied by the aforementioned factor. This property enables a good determination and subtraction of wavelength dependent features at a given position, and allows the detection of physical objects in the vicinity of the star.

In the case of long slit spectroscopy, an infinity of images of the slit at increasing wavelengths are dispersed and superimposed to create a spectrum such as that shown in Fig. \ref{figure:typical_spectrum}. The wavelength-dependent artifacts such as speckles create oblique lines\footnote{In the case of a spectrum with constant spectral interval (case considered in this work), the wavelength dependent artifacts create straight oblique lines. In a more general case where the spectral interval may vary (constant resolution for instance), the artifacts follow a curve as a function of wavelength.} in the spectrum. A bright speckle will see its size increased with wavelength, creating a bright line in the spectrum with increasing width. A physical object creates a straight line at a fixed radial separation from the star, and its FWHM increases linearly with wavelength. Our method uses these geometrical properties to separate the planet and the star spectra, as proposed by \citet{thatte2007} with spectral deconvolution (SD) in the case of IFS observations.

\subsection{Method description}
\label{section:method_description}

\subsubsection{General concept}
\label{section:general_concept}

In the image representing the spectrum, each column of pixels corresponds to a spectral interval $\Delta\lambda$ of a few nanometers, centered on a particular wavelength. From now on, we assume that a given column of pixel $i$ in the spectrum corresponds to a single wavelength equal to the central wavelength $\lambda_{i}$ of the spectral interval $\Delta\lambda_{i}$. This approximation is valid as long as $\Delta\lambda_{i}/\lambda_{i} \ll 1$, which is the case for the wavelength range and spectral interval considered in our work (see Sect. \ref{section:lss_simulations}) where $\Delta\lambda_{i}/\lambda_{i} < 0.015$ in any case.

\begin{figure}[pt]
  \centering
  \includegraphics[width=8cm]{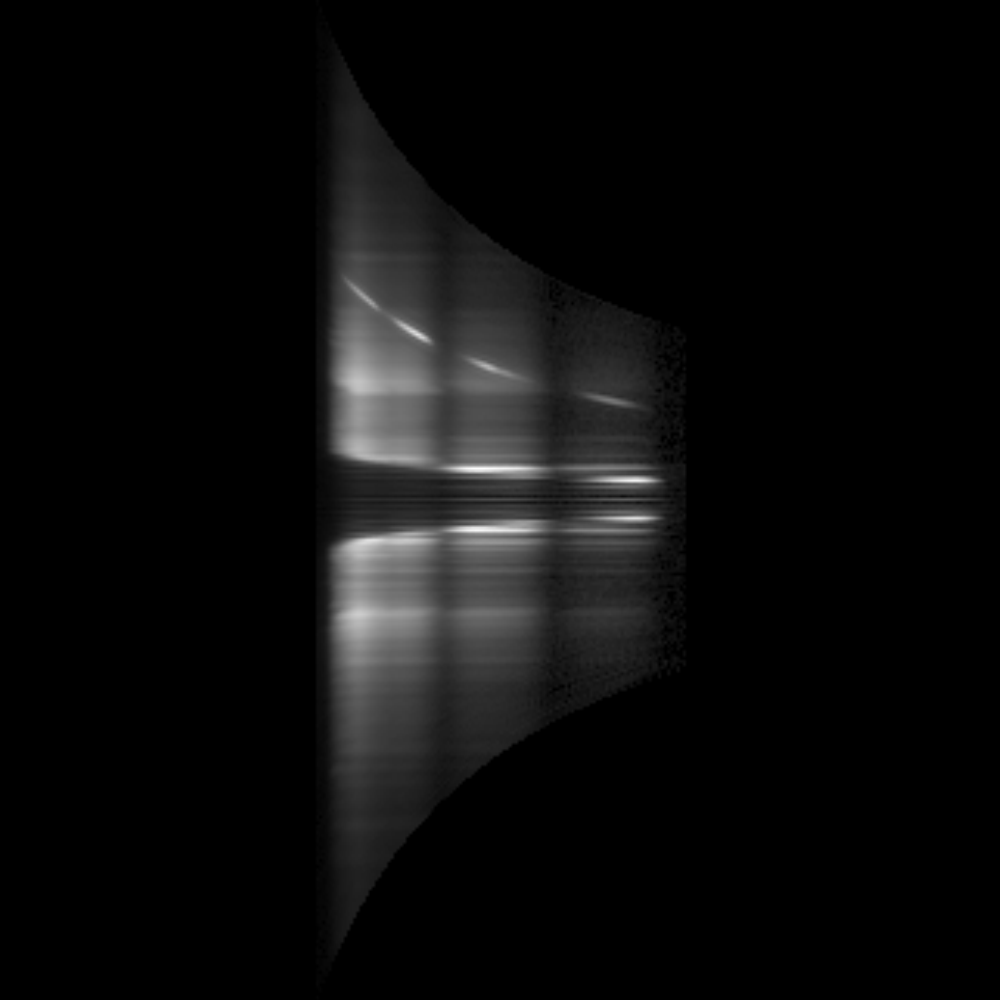}
  \caption{Same spectrum as Fig. \ref{figure:typical_spectrum} where each column has been rescaled according to its corresponding central wavelength to correct for the spectral dependence of the speckles. The speckles are now along straight horizontal lines, and the planet follows a $1/\lambda$ law. The displaying scale is the same as in Fig. \ref{figure:typical_spectrum}}
  \label{figure:rescaled_spectrum}
\end{figure}

The first step corrects the spectral dependency of the speckles by spatially rescaling each column of pixels by a factor $\alpha_{i}=\lambda_{0}/\lambda_{i}$, where $\lambda_{0}$ is the shortest spectrum wavelength. Since $\alpha_{i}$ is not an integer for most columns, each column was resampled in the spatial direction to rescale its length from one integer number of pixels to another integer number of pixels. The number of resolution elements is fixed by the instrumental setup (i.e. the platescale) in the spatial direction, and by the dispersive element in the spectral direction. In fact, the spectrum is resampled within a grid $\sim$15 times smaller than the original pixel grid before rescaling the columns by the factor $\alpha_{i}$. When calculating the new integer number of pixels for column $i$, this spatial oversampling allows us to reduce the round-off error introduced by the non-integer factor $\alpha_{i}$. With the approximation mentioned in the previous paragraph, neglecting the spectral interval $\Delta\lambda_{i}$ translates into an error of less than 1.5\% in $\alpha_{i}$. The final rescaled spectrum is shown in Fig. \ref{figure:rescaled_spectrum}.

The rescaling corrects the spectral dependence of each speckle, ensuring that they follow horizontal straight lines, while the companion spectrum follows a $1/\lambda$ law. Speckles are induced by random phase errors in the incoming wavefront that scatters the starlight. The fraction of light scattered in each speckle coming from the companion is negligible compared to the scattered light of the star, so it is safe to assume that the spectral decomposition of a speckle reflects only the star spectrum, multiplied by atmospheric and instrumental transmission. The only difference between one angular position and another is the amplitude of the modulation created by the succession of bright and dark speckles superimposed on the star halo and the Airy pattern. As a result, the star spectrum can be recreated precisely by averaging all the lines for which a full spectrum is available. If the spectrum extends to an angular separation $\pm\rho$~rad from the star, the lines containing a complete spectrum are those with a separation of $\pm\rho D/\lambda_{1}$ cycles per pupil diameter, where $\lambda_{1}$ is the longest wavelength in the spectrum. In the case of coronagraphic LSS, we have to exclude the central lines corresponding to the position of the Lyot coronagraphic mask. We finally obtain a linear average spectrum of the star, which we refer to as the model spectrum.

The next step is to estimate the exact amount of starlight at each angular separation from the star. To do so, we extract a linear spectrum at each angular position at which we fit our model spectrum in amplitude using a least-squares estimation. At the beginning, both the linear and the model spectra are normalized to unity, and we allow our model normalization to vary between a factor of 0.5 and 3 times the initial normalization, in steps of 1\%, during the fitting process. This allows us to optimize the fit of the model to the linear spectrum. Our least-squares statistic is defined to be

\begin{equation}
 s^{2} \equiv \frac{\sum_{i=1}^{N}\left[\beta f_{i}^{\mathrm{model}}-
 f_{i}^{\mathrm{linear}}\right]^{2}}{N},
\end{equation}

\noindent where $f_{i}^{\mathrm{model}}$ and $f_{i}^{\mathrm{linear}}$ are the model flux and the flux of the linear extracted spectrum, at the considered angular separation in each wavelength bin $i$; $\beta$ is the varying normalization applied to the model, ranging from 0.5 to 3 in steps of 0.01; and $N$ is the total number of wavelength bins. The parameter $s^{2}$ therefore evaluates the average deviation between the model and the linear spectrum. The optimal fit to the data is obtained by finding the normalization $\beta_{m}$ that minimizes $s^2$. When the appropriate normalization $\beta_{m}$ has been determined, the model is multiplied by the product of this factor and the initial factor used to normalize the linear spectrum to unity.

\begin{figure}[pt]
  \centering
  \includegraphics[width=8cm]{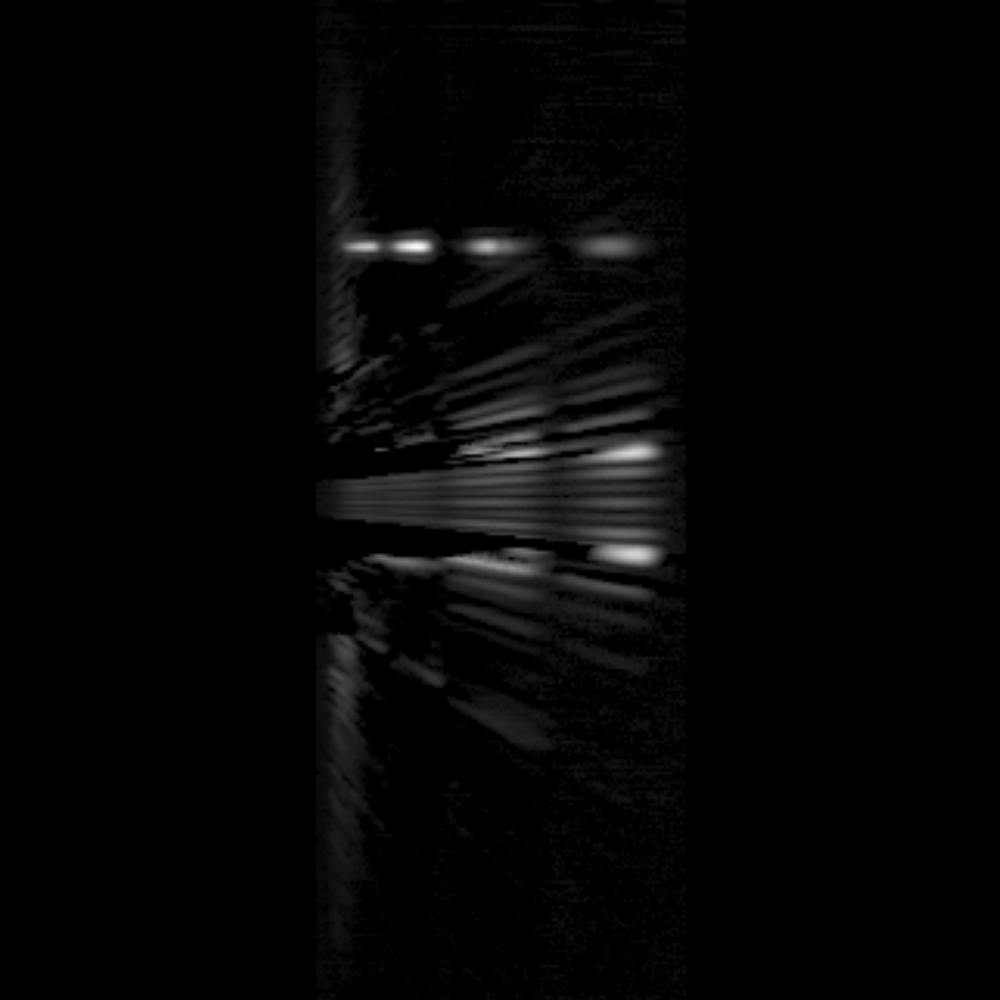}
  \caption{Final spectrum after removing scattered starlight and scaling its dimensions back to their original size. Most of the starlight has been removed, and the companion spectrum can be easily extracted. Some residuals are visible, especially close to the center where the star lies. The displaying scale is the same as in Fig.  \ref{figure:typical_spectrum}}
  \label{figure:corrected_spectrum}
\end{figure}

By doing so for every angular separation in the spectrum, we construct a ``clean'' spectrum i.e. a spectrum devoid of light from the companion, that we will refer to as the reconstructed spectrum from now on. The reconstructed spectrum is subtracted from the rescaled spectrum, to remove most of the scattered starlight and reveal the spectrum of the companion. Finally, each of the resulting spectrum columns needs to be scaled back to its original size to be able to recover the companion spectrum along a straight horizontal line. It is then possible to extract a final spectrum of the companion unaffected by any scattered starlight. The final spectrum is illustrated on Fig. \ref{figure:corrected_spectrum}.

\subsubsection{Discussion and implementation details}
\label{section:discussion_and_implementation_details}

A few important details need to be emphasized to achieve a robust implementation of this data analysis method. First of all, this method is affected by the same conditions about the \emph{bifurcation point} defined by \citet{thatte2007}. The issue is not to estimate the starlight uncontaminated by light from the companion, which is achieved by averaging the spectrum, but to have a sufficient number of points to complete an accurate amplitude fitting. To estimate reliably the amount of scattered starlight at a given angular position, it is necessary that at least one data element that is uncontaminated by the light of the companion. The more data points used to fit the linear average spectrum, the more accurate the estimation of the stellar contribution, and the more reliable the characterization of a companion will be.

Secondly, for very bright companions, a mask is applied to the data being fitted to ensure that the light of the companion does not contaminate the amplitude fit. If a companion is very bright compared to the scattered starlight that we wish to evaluate, the fitting will be completely biased: the reconstructed spectrum will be overestimated and the companion spectrum continuum will be underestimated. When subtracting the reconstructed spectrum from the observed data, part of the companion light is then removed. This issue is important only for very bright companions, whose position can be easily ascertained from the data to create a mask at the appropriate position. The exact size of this mask can be deduced from the instrumental PSF. We choose to consider a size of $2.5\lambda/D$ to completely mask the PSF core (see Sect. \ref{section:planet_mask_size} for a more complete analysis).

Finally, depending on the slit size and spectral range, it is important to
consider that speckles located at the edge of the slit move out of the slit as
the wavelength increases. For the slit sizes (0.09\as and 0.12\as, see
Sect. \ref{section:lss_simulations}) and wavelength range (J, H and K bands)
considered, a significantly large number of speckles move out of the slit
($\sim$50\%). The impact of this effect is identical at all positions in the
slit, so the final influence is negligible on data analysis. However it means
that the measured model spectrum is not equivalent to the true star
spectrum. And that discrepancy between the two increases with wavelength as
speckles move out of the slit. This implies that precise photometry of the star cannot be performed on the model spectrum.

\section{LSS simulations}
\label{section:lss_simulations}

Intensive simulations were completed to simulate realistic spectra for testing the method. We used the complete simulation model of the VLT-SPHERE instrument developed for end-to-end simulations. This model is a diffractive code written in IDL (Interactive Data Language) that is based on the CAOS (Code for Adaptive Optics Systems) problem solving environment \citep{carbillet2004} with a package developed for the SPHERE project \citep{carbillet2008}. We do not to describe in detail the content of this simulation code, but present a global overview of the simulated.

The SPHERE package for CAOS is a diffractive end-to-end simulation code that takes into account multiple sources of aberrations, such as atmospheric residuals, AO correction and optical aberrations. It consists of separate modules, which simulate different parts of the instrument i.e. the extreme AO system SAXO, the optical common path, and the science modules. A specific part of the code is dedicated to LSS simulations with coronagraphy. This code does not simulate temporal variation in the aberrations: only the AO-filtered atmospheric residuals are changed during the simulation. The final output of the code are normalized images of the PSF, the coronagraphic PSF, and slit images at different wavelengths.

A second code was developed to scale these normalized images to the correct photometric values according to the instrument design, the star being observed, the distance, the angular separation, and the \teff of the companion. In particular, the code accounts for the global throughput of the instrument, as well as the atmospheric transmission. The code considers OH lines but not their variability. The spectra used to model the companions in the simulations are the latest synthetic spectra generated by the ENS Lyon group with the PHOENIX code \citep{allard2001}. More precisely, we used the models designated DUSTY-2000, COND-2002, and SETTL \citep[and private communication]{allard2001,allard2003}. Finally, a realistic amount of noise was added depending on the characteristics of the particular next-generation AO imaging instruments i.e. detector noise, sky background and instrumental thermal background. The final output was a two-dimensional spectrum of the star with one or more companions at different angular separations.

\begin{table}[t]
 \caption[]{Simulated long slit spectroscopy modes.}
 \label{table:spectro_modes}
 \centering
 \begin{tabular}{c c c c c}
 \hline\hline
 Mode & Band & Resolution$^{\mathrm{a}}$ & $\Delta\lambda$ & Slit width \\
      &      &            & (nm/pixel)      & (\as)      \\
 \hline
 LRS  & JHK  & 35         & 13.6            & 0.12       \\
 MRS  & J    & 400        & 1.19            & 0.09       \\
 \hline
 \end{tabular}
 \begin{list}{}{}
 \item[$^{\mathrm{a}}$] Resolution is given at $\lambda = 950$~nm.
 \end{list}
\end{table}

We simulated 0.12\as and 0.09\as wide slits with a 0.20\as radius Lyot coronagraph at the center, for an 8 meter diameter telescope. The number of wavelengths and the spectral interval was carefully chosen to produce spectra at low and medium resolution, refered to as LRS and MRS respectively (see Table \ref{table:spectro_modes}). The amount of optical aberrations was set to be $\sim$50~nm RMS before the coronagraph, and $\sim$40~nm RMS after the coronagraph. The atmosphere was simulated by a set of decorrelated phase screens, each corresponding to approximately 80~nm RMS of wavefront aberrations after AO correction. We limited our simulations to 100 phase screens to produce a smooth star halo. In theory, this number is too small to produce a ``true'' long exposure. If we assume that atmospheric residuals have a correlation time of the order of 1~ms and that a typical exposure time for a DIT is 10~seconds, then 10,000 decorrelated phase screens would be required to produce a long exposure. However, in practice using fewer phase screens has proven to give good results in various conditions, while significantly reducing the computing time of the simulations. The total amount of aberrations is $\sim$110~nm RMS, which corresponds to a Strehl ratio of 83\% at 1.6~$\mu$m. Although high compared to existing instruments with conventional AO (typical Strehl ratio for VLT-NACO in standard conditions is between 40\% and 60\% in K band, see \citealt{clenet2004}), these overall performances are realistic for new instruments with high order AO systems and exceptionally high optical quality such as VLT-SPHERE. To generate a significant amount of data we created 6 independent sets of data in LRS with our simulation code by changing the random number generator seeds for every set. Each set produced a spectrum with a unique realization of the random atmospheric and instrumental aberrations. For MRS, only 2 different data sets were generated in J band because of the considerable number of wavelengths that had to be simulated.

\begin{figure*}
 \centering
 \includegraphics[width=18cm]{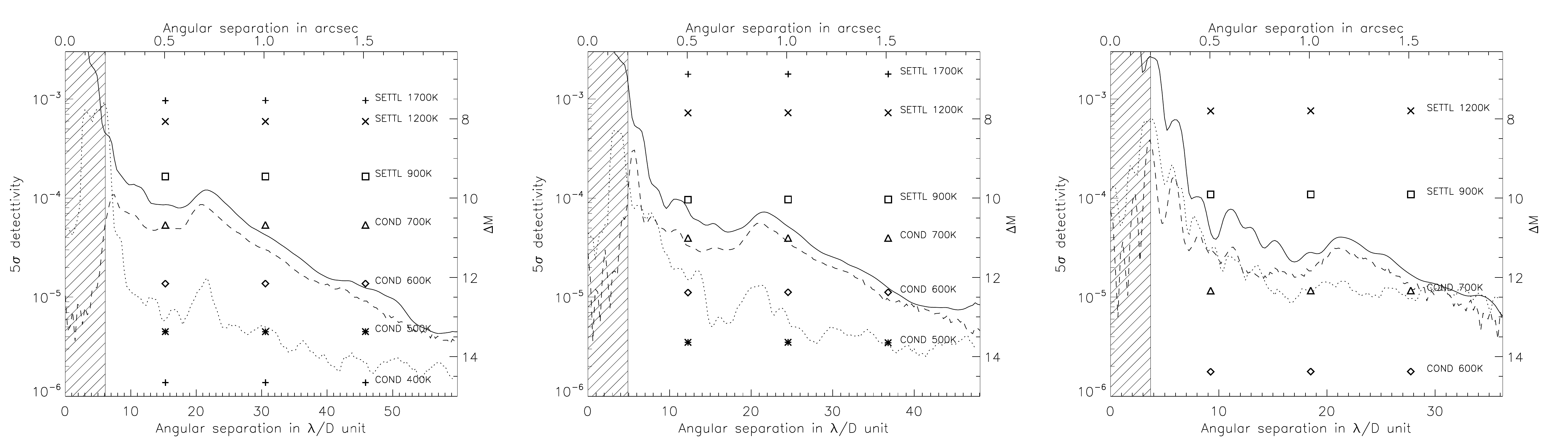}
 \caption{Contrast curves for a M0 star at 10~pc in J band (left, $\lambda  = 1.27$~$\mu$m), H band (middle, $\lambda = 1.58$~$\mu$m), and K band (right, $\lambda = 2.10$~$\mu$m), for an exposure time of 1 hour at LRS. The hatched area is the part covered by the coronagraph mask. The plot shows the non-coronagraphic PSF (plain line), the coronagraphic PSF (dashed line), and the contrast level after applying our method (dotted line). The simulated companions with different \teff are also plotted at 3 angular separations, 0.5\as, 1.0\as and 1.5\as, with their effective temperature and the atmosphere model.}
 \label{figure:contrast_LRS_M0_10pc}
\end{figure*}

Photometry for each data set was calculated to be representative of one hour exposure time for M0 and G0 stars at a distance of 10~pc, and companions with effective temperatures ranging from 400~K to 2700~K at angular separations of 0.5\as, 1.0\as, and 1.5\as (transposing to 5, 10 and 15~A.U. at 10~pc respectively). Table \ref{table:contrast} lists the atmosphere models used, their effective temperatures, and contrast in the JHK bands around M0 and G0 stars. Different noise sources were added as residual variance simulate the detection process realistically i.e. photon noise, flat field noise (0.1\%), and read-out noise (15~e$^{-}$). Thermal background coming both the sky and the instrument were also added: the instrumental thermal background was simulated to be photon noise induced by a constant emission with a fixed number in K band: 344 photon~sec$^{-1}$~pixel$^{-1}$; the sky background has been simulated as a two-dimensional spectrum calculated to match the sky emission at ESO-Paranal observatory in Chile ($86 \times 10^{-16}$~photon~sec$^{-1}$~m$^{-2}$~as$^{-2}$ at $\lambda=2.2$~$\mu$m, values calculated from \citealt{patat2008}). The final output of the simulation represents science calibrated images that were reduced following the usual steps of data reduction.

\section{Results}
\label{section:results}

\subsection{Data analysis performance}
\label{section:data_analysis_performance}

The global performance takes into account two aspects: (a) the achievable contrast, and (b) the fidelity of the extracted companion spectrum with respect to the original spectrum. We discuss our results in terms of these complementary aspects in the following sections.

\subsubsection{Contrast reduction}
\label{section:contrast_reduction}

\begin{figure}[t]
 \centering
 \includegraphics[width=8cm]{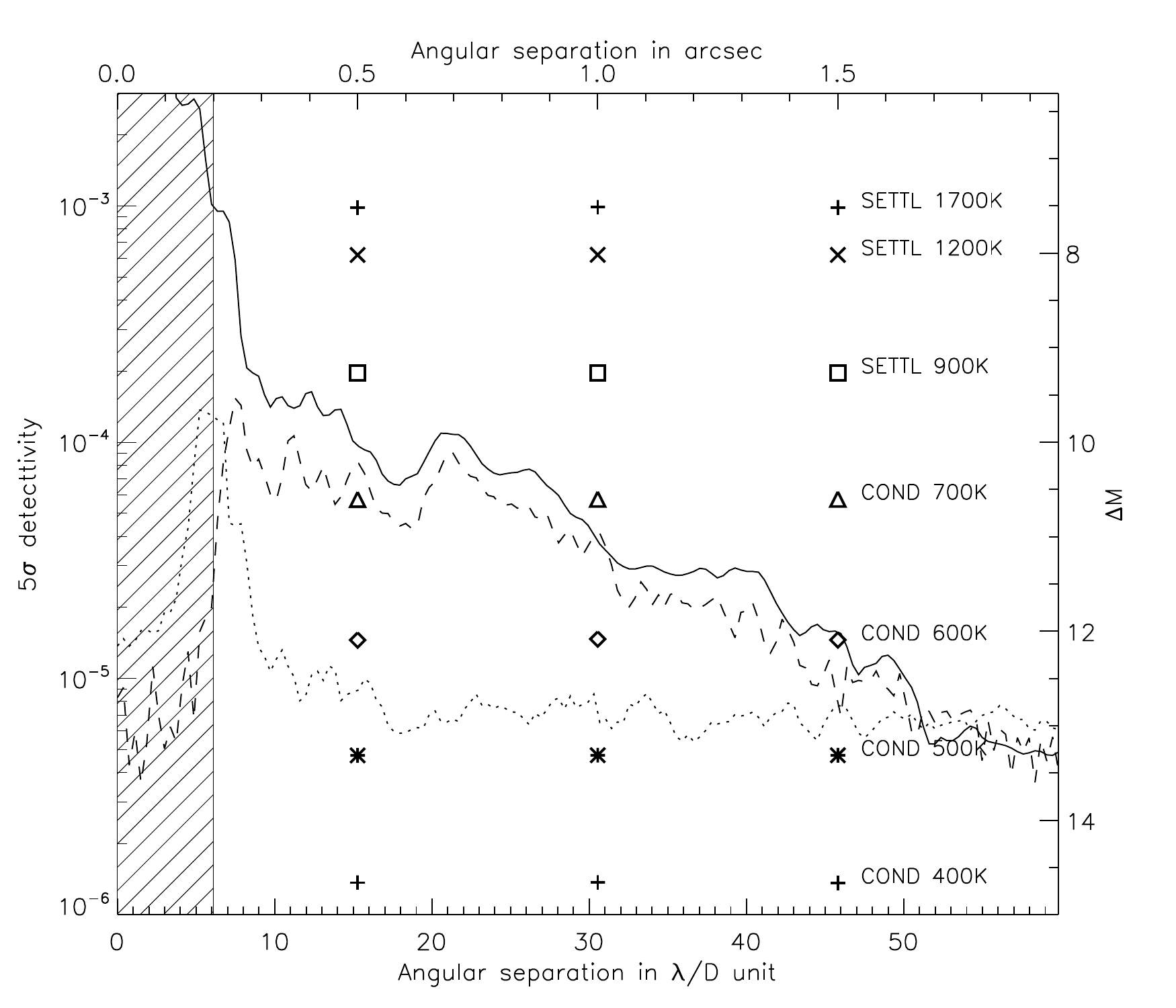}
 \caption{Contrast curves for a M0 star at 10~pc in J band ($\lambda = 1.27$~$\mu$m), for an exposure time of 1 hour at MRS. The hatched area is the part covered by the coronagraph mask. The plot shows the non-coronagraphic PSF (plain line), the coronagraphic PSF (dashed line) and the contrast level after applying our method (dotted line). The simulated companions with different \teff are also plotted at 3 angular separations, 0.5\as, 1.0\as, and 1.5\as, with their effective temperature and the atmosphere model.}
 \label{figure:contrast_MRS_M0_10pc}
\end{figure}

\begin{figure*}
 \centering
 \includegraphics[width=18cm]{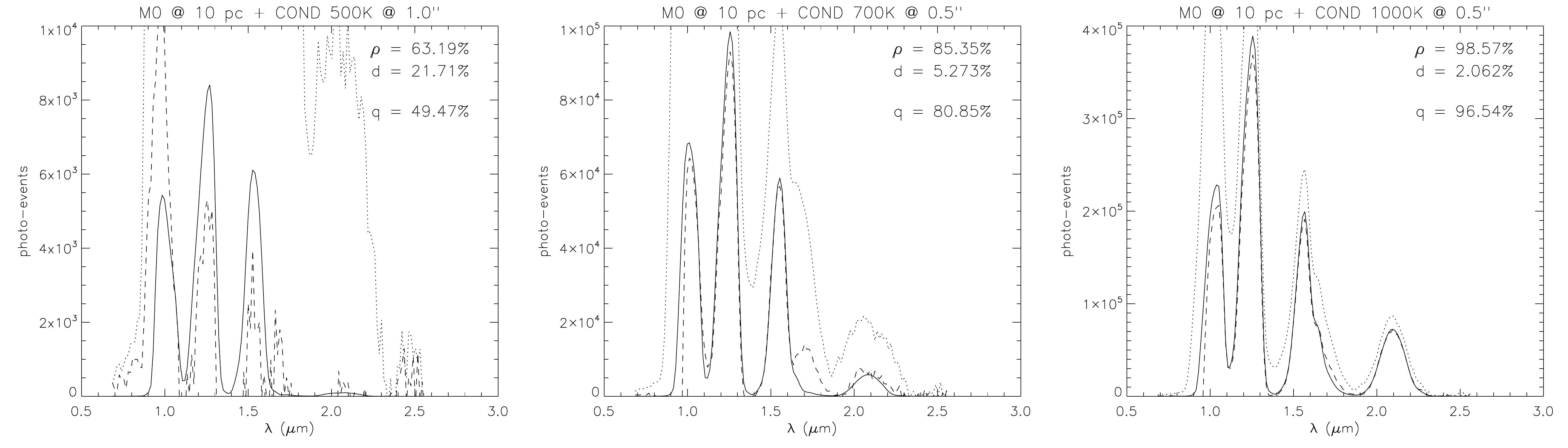}
 \caption{Example of quality factor for 3 different companions around an M0 star at 10~pc in LRS. In the upper-right part of each plot, we indicate the correlation factor $\rho$, the discrepancy factor d, and the final quality factor q. The plot shows the spectrum that was introduced as input in the data analysis (planet only, plain line), the spectrum before data analysis (star  and planet, dotted line), and the spectrum extracted as a result of data analysis (dashed line).}
 \label{figure:quality_example_LRS_M0_10pc}
\end{figure*}

The goal of the data analysis in LSS was to remove the scattered starlight and to improve the contrast. If we consider starlight to be the main noise source, this operation will improve the signal-to-noise ratio (SNR). Figures \ref{figure:contrast_LRS_M0_10pc} and \ref{figure:contrast_MRS_M0_10pc} present the 5$\sigma$ detectivity at different wavelengths for a M0 star at 10~pc and companions of increasing \teff at 3 angular separations, at LRS and MRS respectively. The contrast curve was obtained by measuring the standard deviation of the corrected spectrum in a window about the considered wavelength of a width equal to that of the slit image on the detector. All curves were normalized to the non-coronagraphic PSF.

At LRS, we see a clear improvement in the contrast in J and H bands between the coronagraphic PSF before and after correction. The gain is of the order of 2.5 magnitudes close to the star, and 1 magnitude farther out. Companions as cold as 600~K could be characterized adequately in these bands very close from the star, while for a temperature of 500~K they are detectable at separations larger than 1.0\as. In K band, where the contrast is already favorable, the method is almost inefficient. For the G0 star (not plotted here), the results are similar in terms of magnitude gain. The results are of higher quality in K for the G0 star, where a gain of 1 magnitude is achieved at separations larger than 0.7\as.

At MRS the gain is comparable to the LRS case in J band with a 2.5 magnitude increase in the contrast close to the star, but a constant level is reached at an angular separation of $\sim$0.5\as. For larger separations, the final contrast level remains equal to $10^{-5}$. Further simulations show that this is the noise level of the sky that sets this limit. Using a thinner slit would decrease the sky flux, and then its noise level, probably allowing us to reach performances comparable to the LRS at the same angular separations.

Although calculated for a given instrumental configuration, these contrast plots are representative of our method performance because for a given configuration they show a clear improvement in the contrast. The limiting factor here is probably the correlated speckle residuals that appear on Fig. \ref{figure:corrected_spectrum}. These residuals come from an inaccurate estimation of the scattered light during the amplitude fitting of the model spectrum. Since the model spectrum is an average, it is likely that small variations in the speckle spectrum at each angular separation cannot be measured exactly. 

\begin{table}[t]
 \caption[]{Planetary models included in the simulations.}
 \label{table:contrast}
 \centering
 \begin{tabular}{c c c c}
 \hline\hline
 Model & \teff & Contrast$^{\mathrm{a}}$ M0 & Contrast$^{\mathrm{a}}$ G0 \\
       & (K)   & (Mag)       & (Mag)       \\
 \hline
 COND  & 400   & 15.0        & 17.6        \\
 COND  & 500   & 13.6        & 16.2        \\
 COND  & 600   & 12.4        & 15.0        \\
 COND  & 700   & 11.0        & 13.6        \\
 SETTL & 900   & 9.6         & 12.2        \\
 COND  & 1000  & 9.3         & 11.9        \\
 SETTL & 1200  & 7.5         & 10.0        \\
 SETTL & 1700  & 5.8         & 8.4         \\
 SETTL & 2400  & 3.0         & 5.6         \\
 DUSTY & 2500  & 3.0         & 5.5         \\
 \hline
 \end{tabular}
 \begin{list}{}{}
 \item[$^{\mathrm{a}}$] Contrast is calculted over the J, H and K bands.
 \end{list}
\end{table}

\subsubsection{Extracted spectrum quality}
\label{section:extracted_spectrum_quality}

The quality of the data analysis must be quantified to evaluate the confidence that can be attributed to a given result. Two different aspects of extraction quality need to be monitored: (a) the correlation and (b) the discrepancy between input and output spectra. The former is measured by a simple correlation coefficient

\begin{equation}
 \rho_{S_{i},S_{o}}=\frac{E\left[S_{i}S_{o}\right]-
 E\left[S_{i}\right]E\left[S_{o}\right]}{\sigma_{S_{i}}\sigma_{S_{o}}},
\end{equation}

\noindent where $S_{i}$ is the input spectrum, $S_{o}$ is the output spectrum, and $E\left[S\right]$ and $\sigma_{S}$ denote the mean and standard deviation of a spectrum $S$ respectively. The latter aspect is measured by the factor

\begin{equation}
 d_{S_{i},S_{o}}=E\left[\frac{\left|S_{i}-S_{o}\right|}{Max\left(S_{i}\right)}\right].
\end{equation}

\noindent This factor represents the mean discrepancy between the input spectrum and the spectrum that is extracted during analysis. It is normalized such that it provides a value between 0 and 1; the closer to 0 the value is, the higher is the quality of the data analysis. The final quality factor is obtained by multiplying the two previous factors to measure both the correlation quality and the discrepancy between the input and the output:

\begin{equation}
 q_{S_{i},S_{o}}=\rho_{S_{i},S_{o}}\left(1-d_{S_{i},S_{o}}\right).
\end{equation}

A value of $q = 1.0$ denotes a perfect data analysis (i.e. $S_{i}$ and $S_{o}$ are identical), while smaller values denote lower levels of quality. It is necessary to define a ground value above which the data analysis can be considered successful or adequate.

\begin{figure}
 \centering
 \includegraphics[width=8cm]{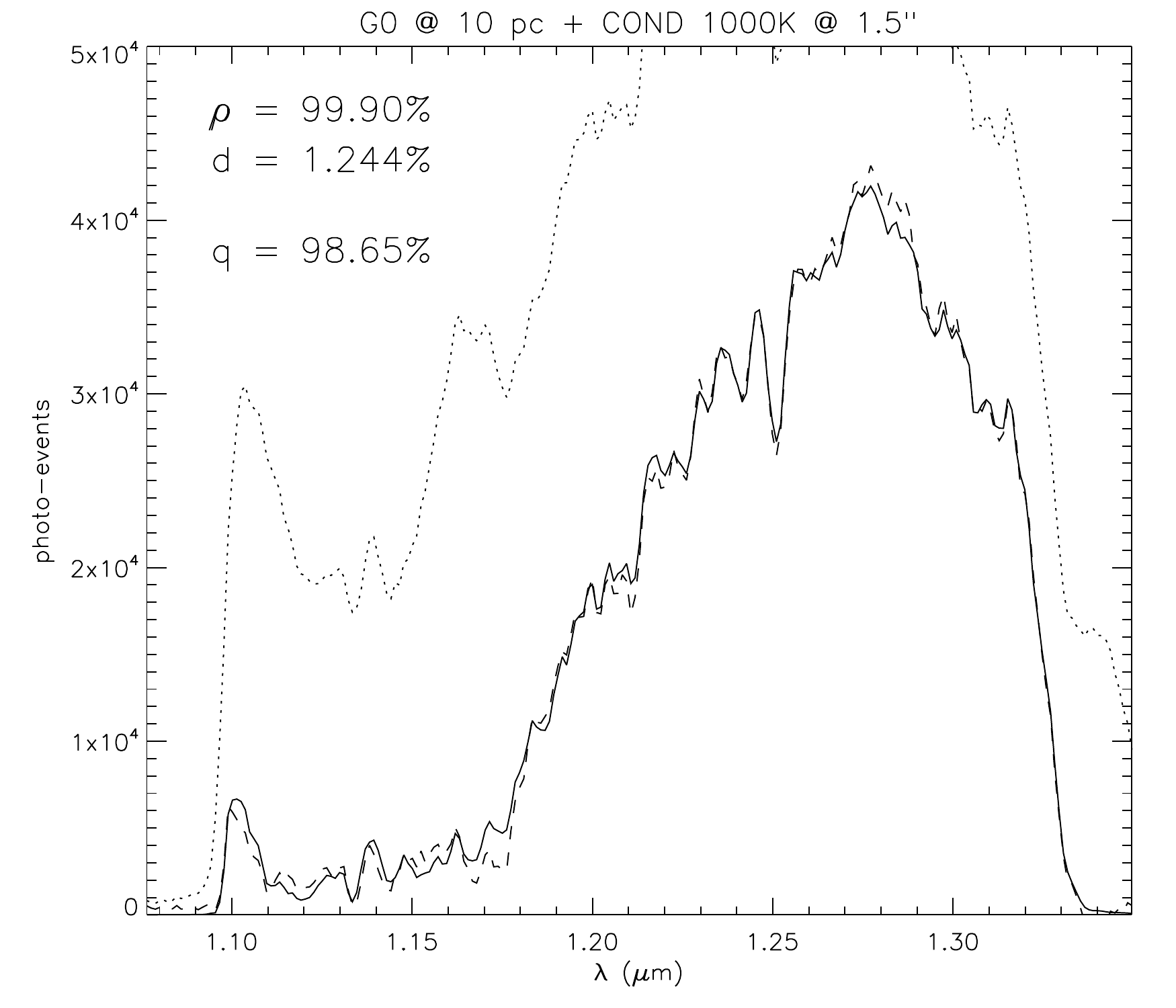}
 \caption{Example of quality factor in J band for an M0 star at 10~pc and a 1000~K companion at 1.0\as in MRS. At this resolution, spectral features are clearly visible. In the upper-right part of the plot is given the correlation factor $\rho$, the discrepancy factor d and the final quality factor q. The plot shows the spectrum that was introduced as input in the data analysis (planet only, plain line), the spectrum before data analysis (star  and planet, dotted line) and the spectrum extracted as a result of data analysis (dashed line).}
 \label{figure:quality_example_MRS_G0_10pc}
\end{figure}

Figure \ref{figure:quality_example_LRS_M0_10pc} presents three cases with very different values of the quality factor at LRS. The lowest quality case corresponds to a 500~K companion at 1.0\as from an M0 star at 10~pc, for which we have a 49.47\% quality factor. The most prominent features in the input spectrum appear in the extracted spectrum, giving an intermediate correlation factor of 63.19\%, but the mean discrepancy between the input and output spectra is quite high, reaching more than 100\% in some areas of the spectrum. Photometry analysis of such a spectrum would produce spurious results. The second case is a 700~K companion at 1.0\as from the same star. The correlation is high, and the mean discrepancy is below 10\%, leading to a high quality factor of 80.85\%. All the main features of the spectrum are reproduced with good accuracy, except for some local residuals that have not been completely eliminated during data analysis. Finally the most successful case is a 1000~K companion at 1.0\as from the same star, which has a quality factor of 96.54\%. All features in the spectrum are reproduced with high precision, allowing an optimal characterization of the companion from its spectrum. The quality factor for a given star clearly depends on the companion \teff and angular separation, i.e. on the overall contrast.

Figure \ref{figure:quality_example_MRS_G0_10pc} shows a good case of data analysis at MRS around a G0 star at 10~pc. The discrepancy is low (less than 2\%) and the correlation almost 100\%, providing a final quality factor of more than 98\%. The dotted line that represents the spectrum before data analysis clearly indicates the gain provided by our method: all scattered starlight has been removed, and the spectrum shows distinct spectral features that were lost in the starlight.

\begin{table}[t]
  \caption[]{Error on \teff estimation as a function of quality factor.}
  \label{table:quality_factor_error_teff}
  \centering
  \begin{tabular}{c c}
    \hline\hline
    Quality factor & \teff error \\
    (\%)           & (K)         \\
    \hline
    0-9            & 1400        \\
    10-19          & 800         \\
    20-29          & 600         \\
    30-39          & 1100        \\
    40-49          & 200         \\
    50-59          & 300         \\
    60-69          & 200         \\
    70-79          & 100         \\
    80-89          & 100         \\
    90-100         & 100         \\
    \hline
  \end{tabular}
\end{table}

One of the primary goals of using LSS on planetary companions is to determine accurately the \teff of the target. We studied the error in the \teff estimation as a function of the quality factor value. To estimate this error, we recalculated photometry with the COND model from the ENS Lyon group for effective temperatures ranging from 400~K to 1500~K in steps of 100~K. Each of these COND models were introduced into our data analysis code for M0 and G0 stars at 10~pc and angular separations of 0.5\as, 1.0\as, and 1.5\as. The output spectra of the data analysis was then compared with the complete COND library\footnote{The complete library contains models with \teff from 100~K to 3000~K in steps of 100~K. It is available at ftp://ftp.ens-lyon.fr/pub/users/CRAL/fallard/} to estimate the \teff from the extracted spectra. Table \ref{table:quality_factor_error_teff} provides the absolute value of the error in \teff as a function of the quality factor (in bins of 10\%). For small quality factors (below 40\%), the error in \teff is significant. Results with quality factors in the range 0 -- 40\% should therefore not be trusted. For quality factors in the range 40 -- 70\%, the error is $\sim$200~K, which represents an error of more than 25\% for a companion with \teff$=700$~K. For a accurate estimation of \teff, quality factors of more than 70\% are required. Above this limit, the error is 100~K or lower. To achieve yet higher accuracy, a finer model grid in temperature should be used. In the following, we choose to set the lower limit for the quality factor to be 80\%, i.e. we consider that results of quality factors smaller than 80\% do not provide a sufficiently precise estimation of \teff.

\begin{figure}
 \centering
 \includegraphics[width=8cm]{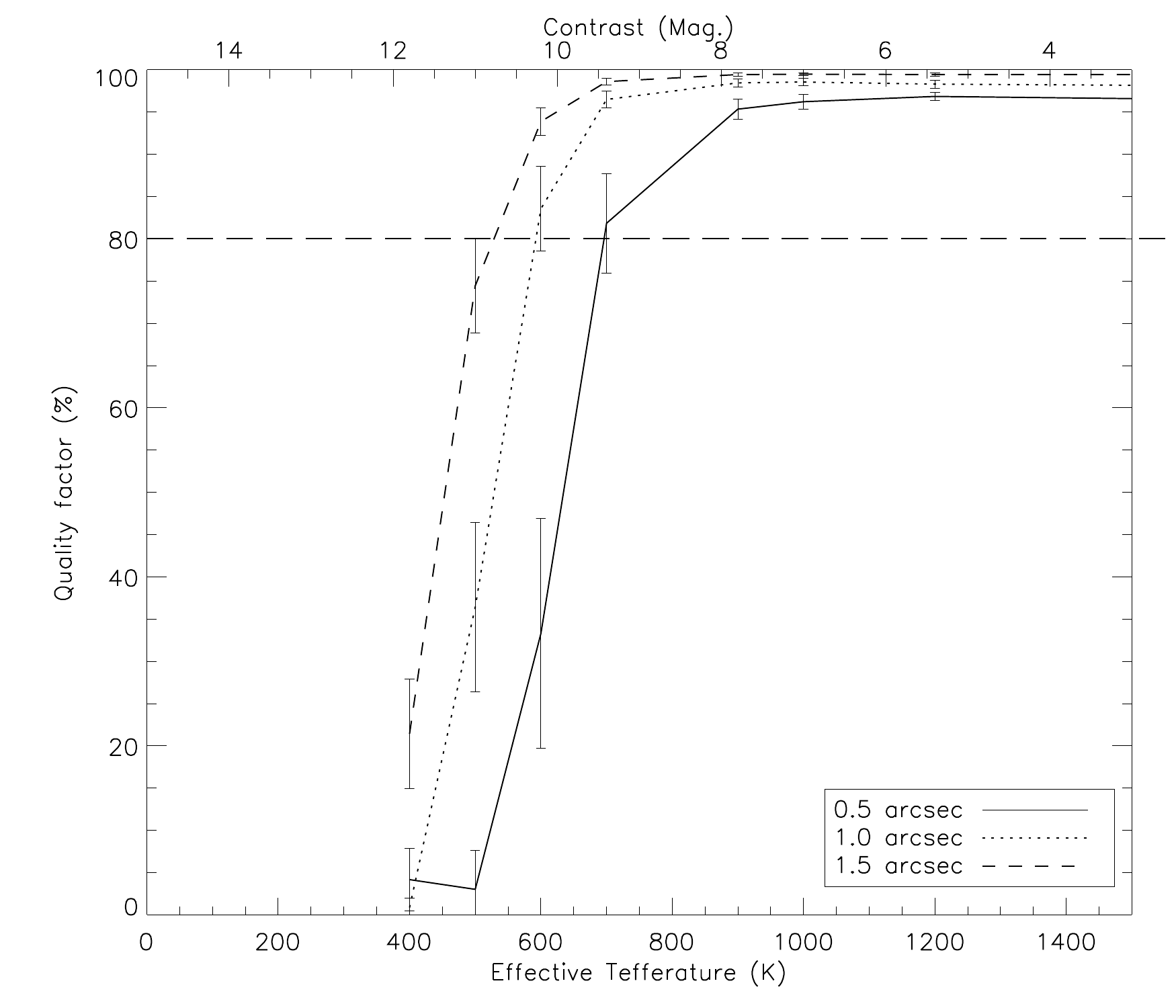}
 \caption{Quality factor for companions around an M0 star at 10~pc in LRS with different \teff at 3 angular separations, 0.5\as (plain line), 1.0\as (dot line) and 1.5\as (dashed line). The horizontal dashed line marks the  80\% level above which the estimation of the \teff can be accurately determined. Error bars represent the standard deviation of the quality factor over the 6 independent data sets that we have simulated in LRS}
 \label{figure:quality_factor_M0_10pc}
\end{figure}

Figure \ref{figure:quality_factor_M0_10pc} indicates the quality factor values for companions with various effective temperatures at 3 angular separations around an M0 star at 10~pc in LRS. The plot gives an overview of the evolution of the quality factor with effective temperature and angular separation. The evolution with \teff is globally the same at each separation: the quality factor slope is very high over a small range of effective temperature, going from close to 0 to approximately 80\%. Above this limit, the slope slowly decreases to reach an almost flat regime above 95\%. The difference with angular separation is the temperature at which the 80\% level is reached: there is a large improvement in contrast between 0.5\as and 1.5\as. At 0.5\as, the 80\% level is reached for a \teff of 700~K (contrast of 11.0 magnitudes), while at 1.5\as the same level is reached for a \teff of 500~K (contrast of 13.6 magnitudes). Results are of even higher contrast for the G0 star (not plotted here), generating wider error bars, in particular for low \teff for which the contrast is very high (15 magnitudes for a 600~K companion). The slope of the curves is different for each angular separation: it is steep for a separation of 1.5\as, and decreases for increasing separations. The curves for separations of 1.5\as, 1.0\as, and 0.5\as reach the 80\% level at different effective temperatures of 700~K, 900~K, and 1200~K, corresponding to contrasts of 13.6, 12.2, and 10.0 magnitudes respectively. This is consistent with the contrast values at which the 80\% level is reached for the M0 star.

\begin{figure}
 \centering
 \includegraphics[width=8cm]{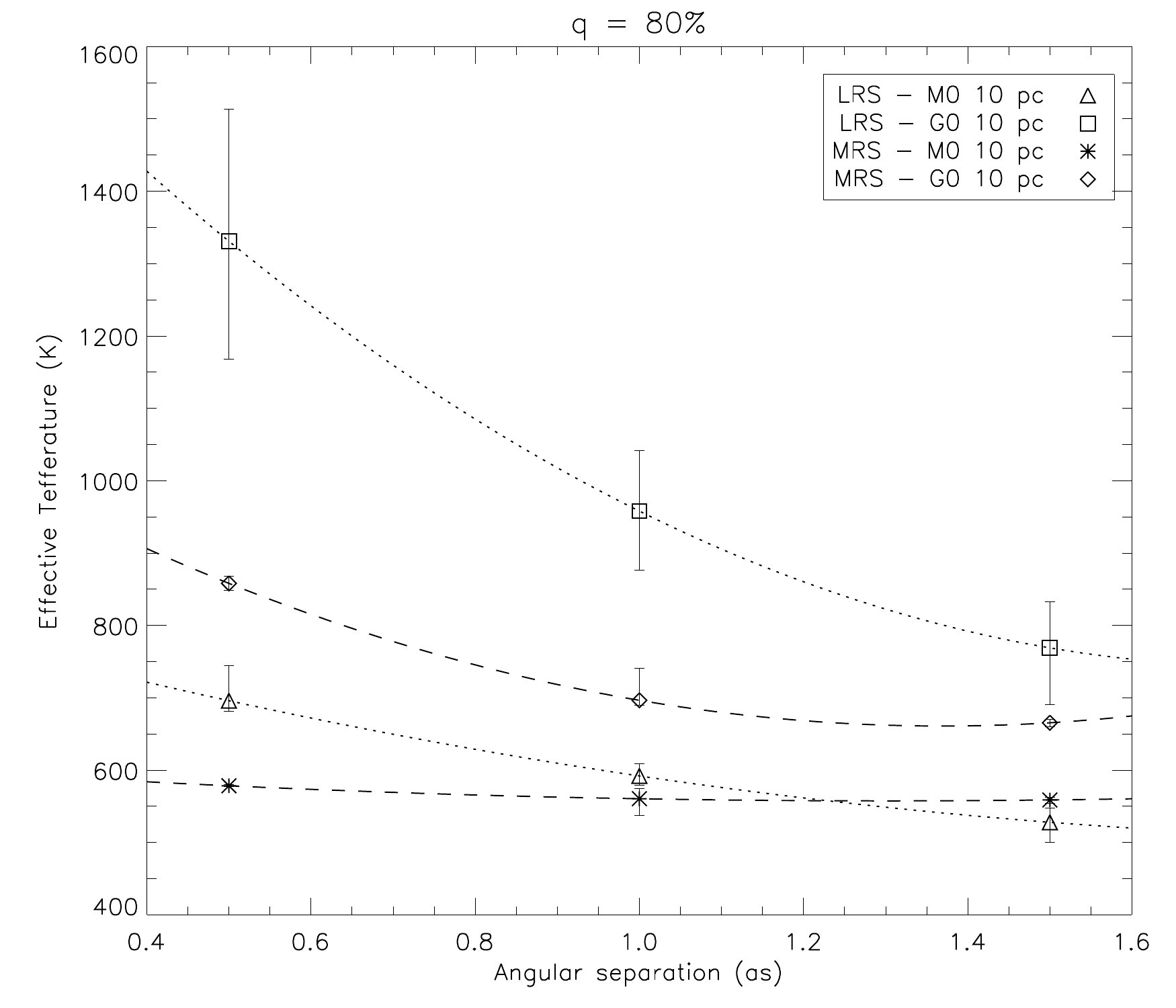}
 \caption{Effective temperature at which a value of 80\% for the quality factor is reached in the case of M0 and G0 stars at 10~pc, as a function of angular separation, in LRS and MRS. A square law has been fitted on the points (dotted lines for LRS, dashed line for MRS) to show the overall evolution with angular separation. The MRS results are only representative  of the J band, and may change slightly for a larger wavelength range.
 }
 \label{figure:quality_summary_M0_G0_10pc}
\end{figure}

Figure \ref{figure:quality_summary_M0_G0_10pc} summarizes the results of the quality factor for both M0 and G0 stars at 10~pc at both LRS and MRS. It shows the effective temperature at which a value of $q = 80\%$ is reached as a function of angular separation. We clearly see the difference in regime between the two stars. A square law was fitted to the data to provide a general idea of the overall evolution, but more points are necessary to have a clearer idea of the real dependence on angular separation.

For the M0 stars, the 80\% level is reached over a small range in \teff (less than 200~K). The influence of angular separation is then quite small, especially at MRS for which the curve is almost flat. This is because at MRS we are limited by the sky level for an M0 star at 10~pc. For G0 stars, the influence of angular separation is far more significant: the range of \teff covered to reach $q = 80\%$ is more than 600~K in LRS. At MRS, the range is only $\sim$250~K, but compared to the evolution in M0 stars, this range is rather large. It finally shows that good characterization of companions with \teff of 600~K orbiting at angular separations of 1.0\as around an M0 star at 10~pc can be achieved. For a G0 star at 10~pc, the characterization of companions with a \teff of 900~K is possible at separations of 1.0\as.

\subsection{Influence of varying parameters}
\label{section:influence_of_varying_parameters}

We study the influence of different data analysis parameters and procedures on the quality factor. To compare the different effects, we focus on the case of an M0 star at 10~pc at LRS, with companions of various \teff at an angular separation of 1.0\as. The MRS case was not considered given the fact that only 2 sets of simulation data in J band were available, but the conclusions for LRS would be applicable to MRS, with a possible scaling factor related to the resolution difference. The two parameters that we studied were the planet mask size (Sect. \ref{section:planet_mask_size}) and the wavelength calibration (Sect. \ref{section:wavelength_calibration}).

\subsubsection{Planet mask size}
\label{section:planet_mask_size}

\begin{figure}
 \centering
 \includegraphics[width=8cm]{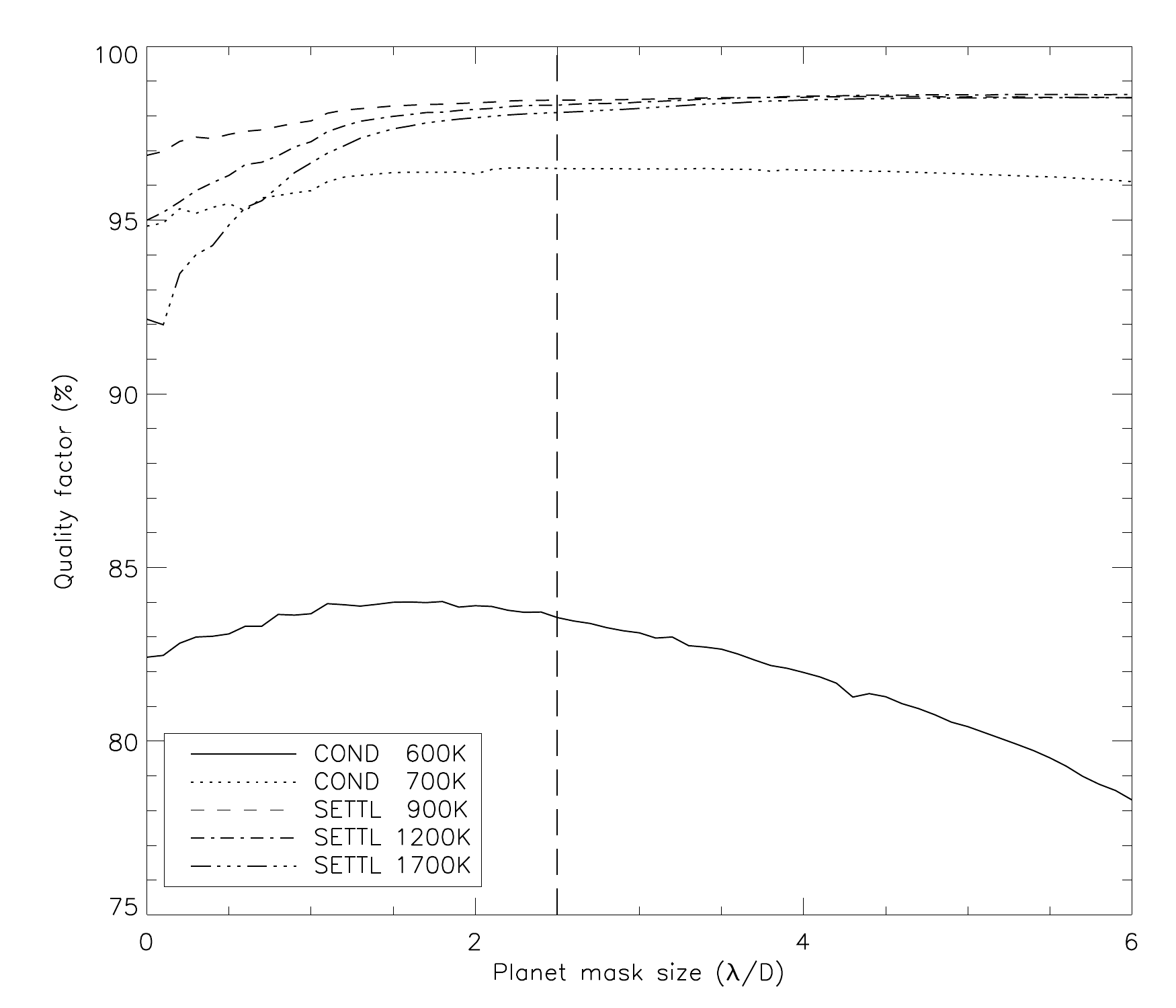}
 \caption{Influence of the planet mask size on the quality factor for 5 companions of increasing \teff orbiting around an M0 star at 10~pc and an angular separation of 1.0\as. Since the diameter of the planet PSF is wavelength dependent, the mask size is calculated in $\lambda/D$ units. The  vertical dashed line marks the $2.5\lambda/D$ limit that was used as default mask size in the simulations presented in this work.}
 \label{figure:influence_mask}
\end{figure}

As explained in Sect. \ref{section:discussion_and_implementation_details}, a mask needs to be applied to the planet signal during the amplitude fitting of the model spectrum to the data to avoid any contribution of the companion signal to the level of scattered light. We studied the effect of changing the mask size as a function of the companion \teff$\!\!$. The results are plotted in Fig. \ref{figure:influence_mask}.

Two distinct regimes can be clearly identified: (i) the case of very faint companions (\teff$\lesssim$~800~K) that cannot be recovered completely during data analysis because of noise and intrinsic limitations of the method, and (ii) the case of fairly bright or bright companions that can be recovered almost completely by the data analysis. In the first case, using a planet mask is unnecessary because it either has no clear effect on the quality factor (COND 700K model) or degrades the quality of the companion signal extraction (COND 600K model). For the 700~K case we see a very small improvement over the 0.0 -- 2.5 $\lambda/D$ range, followed by a slow decrease towards larger sizes. For the 600~K case, a slight amelioration of the quality factor is then followed by a clear degradation of the extraction when mask size increases. In this regime of high contrast where the noise becomes important and may become dominant over the signal, masking part of the signal only reduces the SNR and obviously degrades the data analysis.

For relatively bright companions (\teff$\gtrsim$~800~K), we enter a second regime where the companion signal dominates the noise, and an almost complete recovery of the spectrum is achieveable. In this case, masking the planet signal is absolutely necessary. With a planet mask, the quality factors increase drastically for bright companions: with a mask of size $2.5\lambda/D$, a 6\% gain is observed for the SETTL 1700K model. For colder companions, the effect of a mask of that size is less important, but not negligible. However, above $2.5\lambda/D$, the effect is very small.

The origin of that limit is the encircled energy in the PSF of the companion: with a Strehl ratio of 90\%, more than 77\% of the energy is contained in a circle of diameter $2.5\lambda/D$. In our implementation of the method, we set the size of the planet mask to be this size. In the regime of faint companions, a degradation of $\sim$1\% or less for companions with \teff$\lesssim$~800~K is acceptable compared to the gain of several percent that is obtained for brighter companions. In practice, it could be possible to decide whether or not a mask needs to be used on an individual case basis, but deciding which cases need a mask and which do not is not obvious \emph{a priori}. This is why we decided to use a mask in all cases, without any variation as a function of \teff.

\subsubsection{Wavelength calibration}
\label{section:wavelength_calibration}

Until now we assumed that the wavelength calibration is perfect i.e. that for each pixel column the corresponding wavelength is known precisely. Good wavelength calibration is \emph{a priori} necessary to accurately rescale he spectrum and remove the wavelength dependence, and then be able to characterize the planetary companion. For the considered instrument, the calibration of a long slit spectrograph with a resolution of $R = 35$ can be achieved to a precision of $\sim$0.2\% from the slope of the wavelength dependence, and with a systematic error of less than 15~nm. To understand the consequences of such errors in the calibration more cleary, we simulated two different cases: (i) a case where a systematic error of a few nanometers was added to the calibration; and (ii) a case where the slope of the wavelength calibration was changed.

\begin{figure}
 \centering
 \includegraphics[width=8cm]{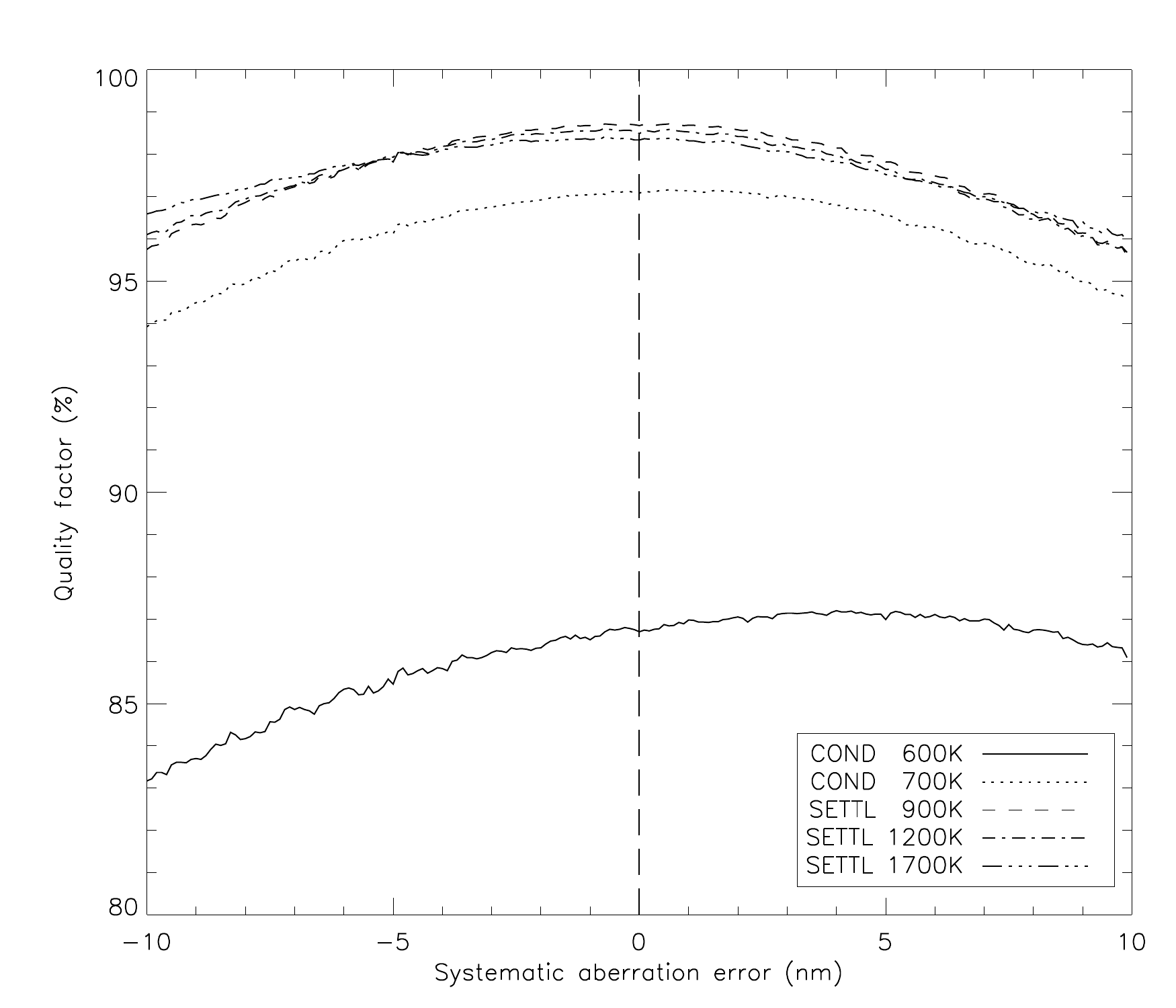}
 \caption{Influence of a systematic calibration error on the quality factor for 5 companions with increasing \teff around an M0 star at 10~pc and an angular separation of 1.0\as. For clarity, the error bars have not been plotted. For each model, they are almost constant: 4.0\%, 0.7\%, 0.25\%, 0.2\% and 0.2\% for the COND 600K, COND 700K, SETTL 900K, SETTL 1200K, and SETTL 1700K models, respectively.}
 \label{figure:influence_systematic}
\end{figure}

Figure \ref{figure:influence_systematic} displays the effect of systematic errors in the calibration for various companions around an M0 star at 10~pc and an angular separation of 1.0\as. Errors of 10~nm were introduced in steps of 0.1~nm, and the resulting quality factor was plotted for various companion \teff$\!\!$. The effect is globally a loss between 2\% and 3\% in the quality factor, which is in any case almost negligible. This does not appear to depend on the \teff of the companion, although the effect is slightly worse for the COND 600K model. The slight loss is attributable to the correlation factor between the input and output spectra that diminishes rapidly when the calibration error increases i.e. the different spectral features are no longer aligned, leading to errors in the characterization of the planet. For the COND 600K atmosphere model, the FWHM of the emission peak in the H band is $\sim$108~nm, so a calibration error of 10~nm produces a $\sim$9\% error in the position of the emission peak in the spectrum.

\begin{figure}
 \centering
 \includegraphics[width=8cm]{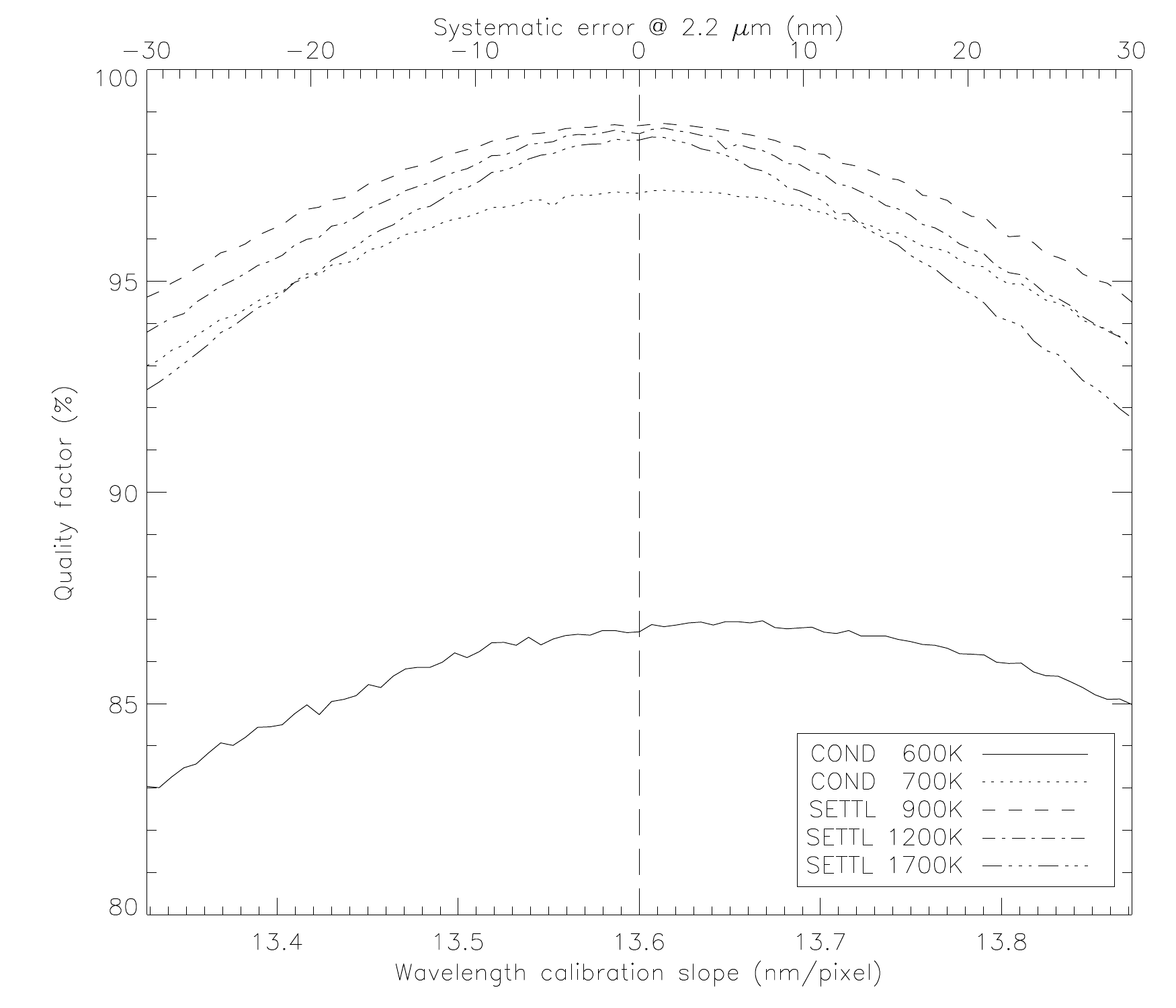}
 \caption{Influence of an erroneous calibration slope on the quality factor for 5 companions with increasing \teff around an M0 star at 10~pc and an angular separation of 1.0\as. The bottom x-axis represents the simulated slope in nm/pixel and the top x-axis represents the resulting systematic error at 2.2~$\mu$m. 13.6~nm/pixel is the optimal slope. For clarity, the error bars have not been plotted. For each model, they are almost constant: 4.1\%, 0.7\%, 0.25\%, 0.23\% and 0.27\% for the COND 600K, COND 700K, SETTL 900K, SETTL 1200K, and SETTL 1700K models, respectively.}
 \label{figure:influence_slope}
\end{figure}

For the second case, we introduced an error in the slope of the wavelength calibration. Figure \ref{figure:influence_slope} illustrates the result of introducing an error of 2\% in the slope, in steps of 0.05\%. In the case that we are considering here, it means varying the slope from 13.328~nm/pixel to 13.872~nm/pixel in steps of $68\times10^{-4}$~nm/pixel. The results are similar to those for the systematic error. The effect is globally the same for all the models, although the two coldest models (600~K and 700~K) appear to be slightly less sensitive to a systematic error than the warmer models. The COND 600K model undergoes a 1\% loss for an error of 0.2~nm/pixel, while the SETTL 1700K model undergoes a 2\% loss for the same error. However, for errors of 0.25\% ($34\times10^{-3}$~nm/pixel) in the slope, the loss is smaller than 0.5\% for all models.

After analyzing the influence of two possible sources of error in the wavelength calibration it appears that there is a clear necessity to achieve the smallest possible systematic error. An error of 10~nm (less than $1\Delta\lambda$) produces losses of several percent in the quality factor, even for bright companions. Reaching an uncertainty level smaller than the spectral interval $\Delta\lambda$ is then an absolute requirement for a proper characterization of planets with our method. The influence of an error in the slope is also important, but the precision that can be achieved in the slope during the calibration procedure is small (0.25\%). At this level of error, the loss in terms of quality factor is negligible.

\section{Conclusions}
\label{section:conclusions}

We have implemented a promising method of characterizing planetary companions using long slit spectroscopy with coronagraphy, which has been considered for very high contrast imaging instruments, such as VLT-SPHERE. The need to develop a specific method was mandatory to remove the scattered light residuals fully from the random speckles pattern in the spectra. Using the linear wavelength dependence of the speckle pattern, we were able to evaluate to high precision the precise contribution of the star to the spectrum, and remove this contribution before extracting a clean planetary companion spectrum. 

The simulations, performed using IDL, allowed us to test our method on realistic data for various cases of contrast in the case of low ($R = 35$) and medium ($R = 400$) resolution spectroscopy with extreme AO and Lyot coronagraphy. The final gain of the method for a 1 hour exposure on M0 and G0 stars at 10~pc is of the order of 0.5 to 2 magnitudes in terms of contrast over the J, H and K bands, compared to the coronagraphic profile. Although not very high in K band, the gain is substantial in J and H, allowing us to study companions with \teff as low as 600~K. Following the evolutionary models of \citet{baraffe2003}, a 600~K companion in a 500~Myr system corresponds to a mass of $\sim$10~$M_{\mathrm{Jup}}$.

To estimate the method efficiency, a quality factor has been introduced to measure the correlation between the input planetary spectrum in the simulation and the extracted spectra, as well as the discrepancy between the two. This allowed us to be confident that our method will allow precise characterization at LRS and MRS of companions as cool as 600~K around M0 stars at 10~pc, or 900~K around G0 stars at 10~pc, for angular separations larger than 1.0\as. For smaller angular separations, results should be extremely valuable for slightly warmer companions.

Finally, we have estimated the influence of two key data analysis parameters at LRS. It is necessary to mask the planet signal during the analysis to avoid overestimating the scattered light level, which would lead to a mis-estimation of the planetary continuum. The method also relies on accurate wavelength calibration to be able to use the known wavelength dependence of the speckles to eliminate them. A systematic error as small as 10~nm can have major consequences on the characterization of the observed objects. 

\vspace{0.5cm}
SPHERE is an instrument designed and built by a consortium consisting of LAOG, MPIA, LAM, LESIA, LUAN, INAF, Observatoire de Gen\`eve, ETH, NOVA, ONERA and ASTRON in collaboration with ESO. 

\bibliographystyle{aa}
\bibliography{paper}

\end{document}